# Conformational Transitions in Silver Nanoparticals: DNA and Photoirradiation

Vasil G. Bregadze[1], Zaza G. Melikishvili[2], Tamar G. Giorgadze[1]

**Abstract**. Photoirradiation of silver nanoparticles in water solution of $NaNO_3$ (0.01M) and in dissolved DNA was investigated by spectrophotometric and thermodynamic kinetic approaches. It is shown that only the irradiated complexes AgNPs-DNA have distinctly expressed isosbestic point. The test with the free AgNPs demonstrates that as a result of photo-irradiation desorption of silver atoms and their oxidation takes place. We have also observed that at photo-irradiation of the complexes by AgNPs-DNA desorption of silver atoms from the surface of AgNPs takes place. Kinetic study of photo-desorption of silver atoms has allowed to estimate desorption rate constant and the heat of adsorption for free AgNPs and AgNPs bound with DNA. On DNA example toxicity of AgNPs at their application to photo-chemo and photo-thermo therapy of cancer is discussed.

[1] Ivane Javakhishvili Tbilisi State University, Andronikashvili Institute of Physics, 6 Tamarashvili Str. 0177 Tbilisi, Georgia;

[2] Georgian Technical University, Institute of Cybernetics. 5, Euli Str., 0186, Tbilisi, Georgia.

**Keywords:** Silver nanoparticles, silver ions, DNA-intercalator complex, optical spectroscopy of intramolecular interactions.

**Corresponding author:** Vasil G. Bregadze vbregadze@gmail.com, v.bregadze@aiphysics.ge

## Introduction

One of interesting tendencies in nano-technology is the application of metal nanoparticles in cancer photo-chemo and photo-thermotherapy. For the purpose it is important to choose materials and sizes of nanoparticles that are capable for: 1) conformational changes; 2) diffusion and 3) chemical transformations in the presence of DNA, light, temperature, ionic strength, redox agents and agents changing environment polarity. A fine example of such material can serve silver nanoparticles (AgNPs) having the size not more than 10 nm [1].

It is hard to overstate the role of metal ions, especially transition ones, in vital activity of organisms. Particular interest causes the interaction of metal ions such as Pt(II), Ag(I), Cu(I) with DNA because metal induced point defects in DNA [2-4] can lead to point mutations and can participate in formation of cross-links between the chains of DNA. One of the interesting examples is the application of cis-diamine -dichlorine- platinium and so-called photo-cis-platinium for tumor treating. More and more articles have been published lately where the usage of metal nanoparticles particularly gold [5-9] and carbon [10,11] ones for tumor photo-thermo-therapy is discussed. Silver nanoparticles (AgNPs) are considerably rarely used for the purpose because one of their substantial properties is stability in solutions.

The goal of the present investigation is to study the interaction of AgNPs with thymus DNA by spectroscopic and thermodynamic methods in 1) darkness; 2) under photo-irradiation and 3) temperature exposure.

**Materials and Methods**

- Colloidal silver suspension in distilled water was prepared of silver nanoparticles (AgNPs) of 1-2 nm size (DDS Inc., D/B/A/, Amino Acid & Botanical Supply).
- Calf thymus DNA produced by Sigma was dissolved in 0.01 M $NaNO_3$ solution – background electrolyte, pH~6.0. DNA concentration was evaluated by UV absorption spectrophotometer. Molecular extinction factor $\acute{\varepsilon}$ = 6600 $cm^{-1}$ $M^{-1}$ (P), $\lambda$ = 260 nm.
- Registration of absorption spectra was carried out by optical fiber spectrometer **AvaSpec ULS 2048 – USB 2.**
- Photoirradiation was carried out in reactor with the fixed light beam in 1cm rectangular fluorescent quartz cell. In the same cell with the interval of 5 min absorption spectra of irradiated solutions were registered by Avantes spectrometer. Before each absorption registration the cell was shut with a shutter that protected the solution from photo irradiation. Registration time was 8 msec. As a source of radiation xenon arch discharge lamp with rating of 35 W in glass balloon was used. At solution irradiation water filter and light filter with light wave transmission $\lambda$=436 nm were used. Radiation power in the cell was 300 mW for water filter and 15 mW for water filter matched with light filter ($\lambda$=436 nm).

**Results and Discussion**

Fig.1 shows the spectra of AgNPs, AgNPs-DNA complexes.

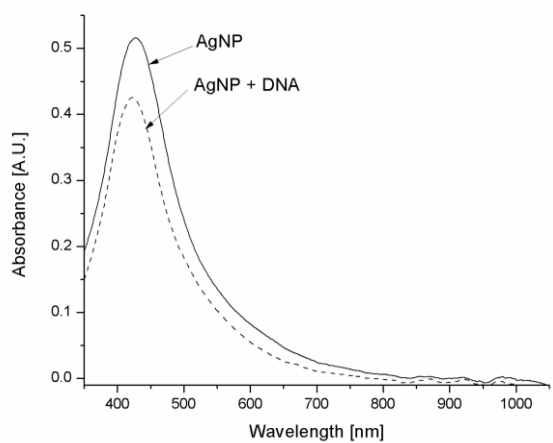

Fig. 1 Absorbtion spectra of AgNPs and AgNPs-DNA complexes.
[AgNPs]-$0.72 \cdot 10^{-4}$ M ($Ag^0$), [DNA]-$1.6 \cdot 10^{-4}$ M(P), [$NaNO_3$]=$10^{-2}$M

The analysis of spectra given in Fig.1 demonstrates that DNA at interaction with AgNPs causes hypsochromic shift of absorption band on 6 nm and 20% hypochromic effect. Blue shift points out that a kind of disintegration (loosening) i.e. attenuation of interactions between silver atoms has taken place. And the decay of intensity means partial corrosion of AgNPs [1].

Kinetics of AgNPs photoirradiation has been studied. Fig. 2 and 3 show superposed absorption spectra of AgNPs and AgNPs in complex with DNA before and after irradiation using water filter.

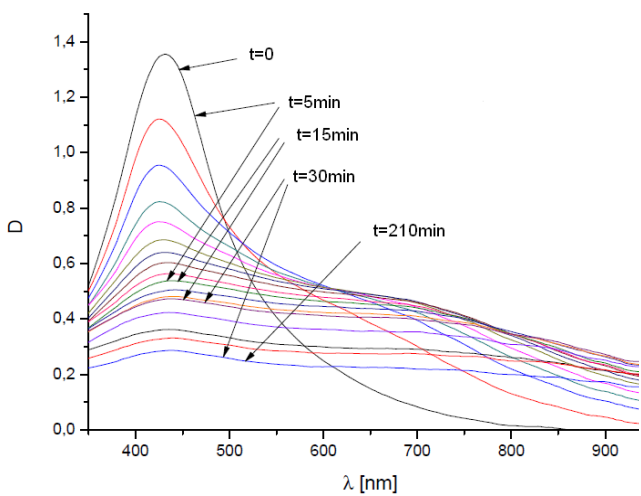
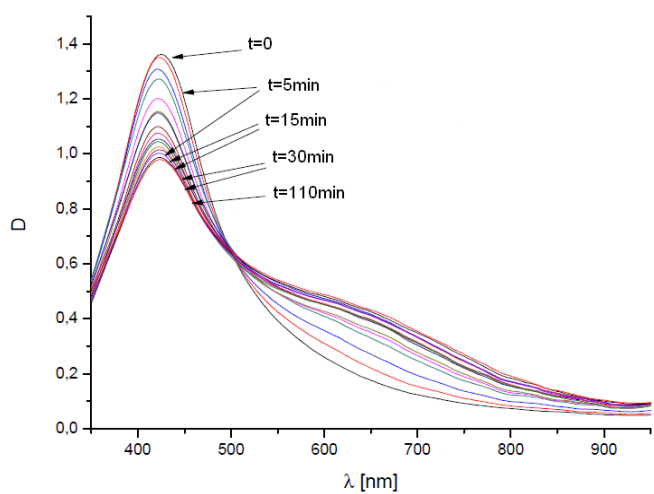

Fig.2 Absorbtion spectra of AgNPs before and after irradiation(5min interval) [AgNPs]-$1.94 \cdot 10^{-4}$ M (Ag$^0$), [NaN0$_3$]=$10^{-2}$M

Fig.3 Absorbtion spectra of AgNPs-DNA before and after irradiation (5min interval) [AgNPs]-$1.94 \cdot 10^{-4}$ M (Ag$^0$), [DNA]-$1.6 \cdot 10^{-4}$ M(P), [NaN0$_3$]=$10^{-2}$M

The analysis of the spectra on Fig. 2 and 3 demonstrate that only the irradiated complexes AgNPs-DNA have distinctly expressed isosbestic point. The test with the free AgNPs shows that as a result of photoirradiation desorption of silver atoms and their oxidation by H$_3$O$^+$ atoms to Ag$^+$ ions takes place . The presence of isosbestic points in the absorption spectra of irradiated AgNPs-DNA complexes proves that the system has two states, i.e. AgNPs-DNA complex has two forms of existence joint by structural photodiffusive transition from one form, e.g. spherical one, to extended long and probably one-dimentional form along DNA double helix. The analysis of the spectra really shows with good correlation ≤5% that the space under the spectra is preserved which means that there are no changes in chromophore electron structure . Besides, half width of absorption spectra $\Delta\lambda_{1/2}$ is changed from 140 nm to 360 mn. Red shift and widening of AgNPs absorption band points out to the typical for molecular systems increase of electron conjugation (linear and cyclic conjugated systems [12]).

4The company nanoComposix [13,14] gives sample absorption spectra for spherical AgNPs with particle sizes of 10, 20, 30, 40, 50, 60, 70, 80, 90 and 100 nm at the same mass concentration 0.02 mg/ml. The given data show that with the increase of the size of the particles widening of the red shift of absorption band can be observed. It is notable that despite the growth of the particle size, i.e. decrease of their total number in the solution, the intensity of absorption bands for nanoparticles with the sizes from 10 to 40 nm is not practically changed, only a small shift of absorption band maximums can be observed. The above explicitly points out that chromophore units are silver atoms and not nanoparticles. Thus, we can draw a conclusion that silver atoms in AgNPs are sufficiently isolated and bound together by dispersion interaction (induced dipole-induced dipole). As these interactions are performed in water surrounding they should be considerably amplified at the expense of so called hydrophobic effect [15], that means compaction and then minimization of the surface (decrease of the system entropy).

We especially point out that in nanoparticle, which consists of one kind of atoms, along with the mentioned dispersial interaction, the so-called resonance interaction should take place [16]. Such types of interactions are typical for molecular crystals and they usually lead to exiton splitting of the principal absorption band. Inevitable condition for exiton splitting is the presence of a system consisting of identical groups and having hard structure [17]. The absence of splitting can mean that AgNPs under investigation (fig.1) have liquid structure resembling a drop which under definite conditions (such as temperature, photo-irradiation, variations in dielectric constant of the environment) should be characterized by conformational transitions. For the estimation of the point we have carried out thermodynamic kinetic analysic of absorbtion spectra of AgNPs (see fig.2 and 3).

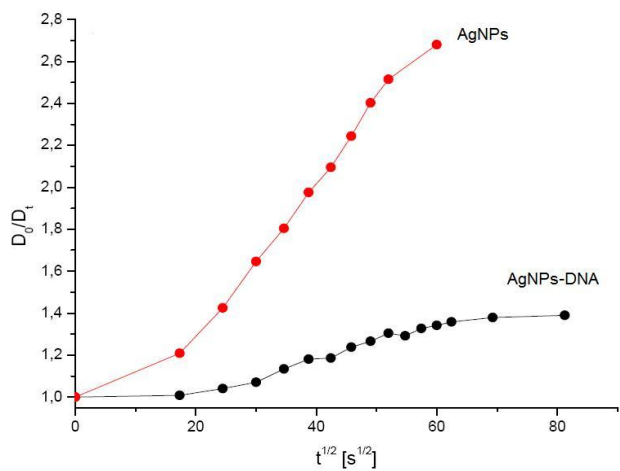

Fig.4 Kinetic curves of photodesorbtion in $M_t/M_e$ and $t^{1/2}$ coordinates for free AgNPs and AgNPs bound with DNA

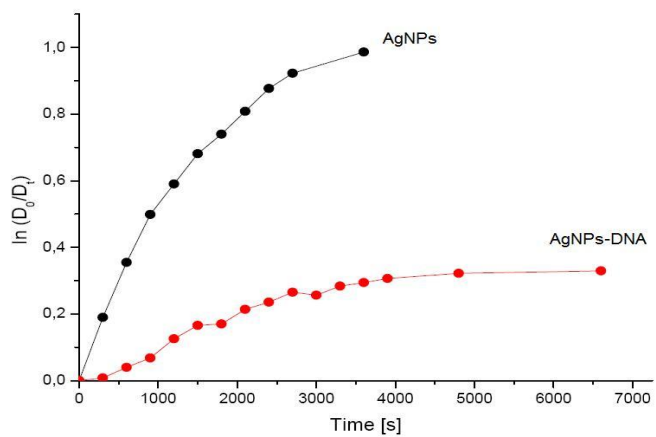

Fig.5 Kinetic curves of photodesorbtion in $\ln[M_e/(M_e - M_t)]$ and t coordinates for free AgNPs and AgNPs bound with DNA

Along with spectroscopic analysis of the curves given on Fig. 2 and 3 thermodynamic kinetic analysis of the mechanism of AgNPs photo diffusion on DNA double helix has been carried out. The mechanism consists of: 1 – desorption of silver atoms from the surface of nanoparticles bound to DNA; 2 – adsorption of silver atoms on DNA surface and 3 – creation of cross-links between silver atoms and DNA chains.

Let's consider the changes in absorption spectra for photo-irradiated free AgNPs and AgNPs–DNA complex given in Fig. 2 and 3 versus the duration of irradiation in $M_t/M_e$ and $t^{1/2}$ (see Fig.4). $M_e$ is the number of silver atoms in nanoparticles at the beginning (adsorption $A_t$ when $\lambda=430$ nm at t=0), $M_t$ is molar quantity of silver atoms desorbed by the time moment t (difference between absorption $A_{t=0} - A_t$ at $\lambda=430$ nm). As it can be seen the curves in Fig.4 have S-shape form both for photo desorption kinetics of silver atoms from the surface of free AgNPs and AgNPs-DNA complexes. S –shape appearance of the curves denotes that photo-induced desorption of atoms is a complex and multiphase process [18]; it is diffusion of silver atoms from the inner part of a nanoparticle to its surface, conformational changes in the particles especially in those ones that are adsorbed on DNA surface. Next we consider the results given in Fig. 2 and 3 in $ln[M_e/(M_e - M_t)]$ and t coordinates, which are given in Fig 5.

The analysis of the curves shows that only initial stage of the given curves of desorption kinetics obey linear law of first-order equation $ln[M_e/(M_e–M_t)]=kt$. The constants of desorption rate of silver atoms from the surface of AgNPs has been evaluated from the slopes of the curves and the date are: for free AgNPs $k_d = (5.5\pm0.2)\cdot10^{-4}s^{-1}$. The values allow us to evaluate desorption reaction activation energy $E_d$ using the equation $k_d = \upsilon_0 exp(^{-E_d}/_{RT})$, where $\upsilon_0$ is pre-exponential factor assumed as $\upsilon_0 =\sim10^{-10}s^{-1}$ (reciprocal quantity to silver atom oscillation time in nanoparticles). In this case we have got the following values for $E_d$ at T= 300° K: $E_d=\sim18.2$ kcal/mol $Ag^0$ for free AgNPs and $E_d =\sim19.3$ kcal/mol $Ag^0$ for AgNPs bound with DNA. As $E_d=E_a+Q_a$, where $E_a$ and $Q_a$ are activation energies for starting activation and heating of nanoparticles, so $Q_1=18.2$ kcal/mol and $Q_2=19.3$ kcal/mol under the condition that formation of nanoparticles is not an activated value. The values of heat are specific for cluster nanostructures [19,20].

Next we give evaluation of life time for $Ag^0$ complex with DNA. As far as in 1996 one of the authors [21] proposed a thermodynamic model of interaction between small ligands and DNA. Based on the example of interaction between the ions of transition metals and DNA it was shown that the life time of the complexes $\tau$ is connected to equilibrium characteristic by stability constant K and equation $\tau=\tau_0 xK$, where $\tau_0$ is the duration of the fluctuation excitation of the adsorbing ligands or molecules interacting with a solid surface and it lyes between $10^{-11}$ and $10^{-10}$ sec. Silver atom photodesorption from the surface of nanoparticles, which are in DNA complex, has desorption activation energy of 19.3 kcal/mol $Ag^0$ and, consequently, we can assume that photoindused diffusion of AgNPs on DNA double helix takes place along with activation of silver atoms desorbtion energy equal to 19.3 kcal/mol $Ag^0$ and thus we can state that the energy of $Ag^0$ interaction with DNA double helix is not less than 19.3 kcal/mol. In accordance with the equation $\Delta G = -RT\ln K$ and with the application of the evaluated energy 19.3 kcal/mol we assume that the stability constant of the complex is not less than $10^{14}$ and, consequently, life time of the complexes is equal to $10^4$ sec. Life-times of 1 sec. order and more are characteristic for inter-strand interactions with the partipipation of transition metal ions, so called cross-links [20]. As far as in 1969 Wilhelm and Daune [22] showed that $Ag^+$ ions form cross-links between DNA chains thus releasing protons bound with $N_1$ guanine and $N_3$ thymine into the solution. We have estimated stability constants of $Cu^+$ and $Ag^+$ ions with DNA which are equal to 10.8 for $Ag^+$ and 14.9 for $Cu^+$. Thereafter the change of free energy for $Ag^+$ is 15kcal/M and 20.5 kcal/M for $Cu^+$ and lifetimes are 0.63 sec for $Ag^+$ and $8.6 \times 10^3$ sec for $Cu^+$ [4].

**Conclusions**

Using spectrophotometry and thermodinamic approaches we have shown that 1) at interaction with DNA-AgNPs are adsorbed on it and only partial corrosion of nanoparticles at the level of $Ag^+$ ions is observed; 2) at photoirradiation ($\lambda$=436 nm or full spectrum of visible band) desorption of silver atoms from the surface of AgNPs takes place. The atoms are first adsorbed on the surface of DNA and then penetrate inside the double helix (cross links between complementary DNA base pairs) making prolate stretched structure (AgNPs absorption spectrum width is changed from 140 nm to 360 nm at half-height); 3) Kinetic study of photo-desorption has made it possible to determine desorption rate constant $K_d$ and adsorption heat $Q_a$ that are equal to $K_d = (5.5\pm 0.2) \times 10^{-4} s^{-1}$ ; $Q_a \geq 18.2$ kcal/M $Ag^0$ for free AgNPs and $K_d = (8.9\pm 0.5) \times 10^{-5} s^{-1}$; $Q_a \geq 19.3$ kcal/M $Ag^0$ for AgNPs bound with DNA. 4) On DNA example toxicity level of AgNPs is shown that can be application to for medical purposes, e.g for photo-chemo and photo-thermo terapy of cancer.

**Acknowledgments:**

The work was partly supported by the Grant No GNSF/ST09_508_2-230.